\documentclass[prl,aps,twocolumn,groupedaddress,showpacs,floats,floatfix,doublespace]{revtex4}
\usepackage[dvips]{graphicx}
\usepackage{epsfig}
\usepackage{url}
\begin{document}

\title{Critical Phenomena in Head-on Collisions of Neutron Stars}

\author{Ke-Jian Jin${}^{(1)}$ and Wai-Mo Suen${}^{(1,2)}$}

\affiliation{${}^{(1)}$McDonnell Center for the Space Sciences,
Department of Physics, Washington University, St.~Louis, Missouri
63130} 
\affiliation{${}^{(2)}$Physics
Department, University of Hong Kong, Hong Kong}

\date{\today}
 
\begin{abstract} 

We found type I critical collapses of compact objects modeled by a polytropic equation of state (EOS) with polytropic index $\Gamma=2$ without the ultra-relativistic assumption.  The object is formed in head-on collisions of neutron stars.  Further we showed that the critical collapse can occur due to a change of the EOS, without fine tuning of initial data.  This opens the possibility that a neutron star like compact object, not just those formed in a collision, may undergo a critical collapse in processes which slowly change the EOS, such as cooling.


\end{abstract}
\maketitle

\paragraph*{\bf Sec.1. Introduction and Motivation.} 

In previous studies \cite{Miller01a,EvansE02} we found that head-on collisions of two neutron stars (NSs) could lead to prompt gravitational collapses, despite a conjecture to the contrary \cite{Shapiro98a}.  Upon collision, two 1.4 $M_{\odot}$ neutron stars with a polytropic equation of state (EOS) could promptly collapse to form a black hole within a dynamical timescale, while NSs with smaller masses may merge to form a single NS \cite{Miller01a}.  Further, in \cite{EvansE02}, we found that prompt collapses could occur even when the sum of the masses of the two colliding NSs is {\it less} than the maximum mass of a single stable equilibrium star, in the case of NSs with the Lattimer-Swesty EOS.  However, we were not able to determine the exact dividing line between the collapse and non-collapse cases, as a reasonably accurate determination would require numerical simulations with a resolution higher than what our 3D numerical code (GR-Astro \cite{GRAstro}) could achieve on computers we have access to.  

In this work we investigate this dividing line.  For the investigation, we have constructed an axisymmetric version of GR-Astro.  GR-Astro-2D enables high resolution simulations of axisymmetric systems.  Details of GR-Astro-2D and the numerical setup will be given in Sec. 2 below.  In Sec. 3, we report on the type I critical phenomena found at the dividing line for NSs with a polytropic EOS of $\Gamma=2$.    (Critical phenomena in gravitational collapse was first discovered by Choptuik \cite{Choptuik93}; for review see e.g. \cite{Gundlach03,Wang01}).  In the super-critical regime, the merged object collapses promptly to form a black hole, even though its mass could be {\it less} than the maximum stable mass of one single NS in equilibrium with the same EOS. 
In the sub-critical regime, an oscillating NS is formed.   Near the dividing line, the near critical solutions oscillate for a long time, before collapsing or becoming an oscillating NS.  

The exact critical solution presumably will oscillate forever, while neighboring configurations would eventually deviate from the critical solution.  In Sec. 4, we determine the critical index $\gamma$ as the time scale of growth of the unstable mode bringing a near critical solution away from the critical solution.  We found $\gamma = 10.78 (\pm 0.6) M_{\odot}$.  This corresponds to a growth time of the unstable mode of about 0.054ms.  In Sec. 5, we investigate the universality of the phenomena with different critical parameter choices.  Particularly interesting is the case in which the parameter is taken to be the polytropic index $\Gamma$, instead of a parameter associated with the initial configurations.  The same critical index is found.
In Sec. 6, we study gravitational waves emitted by the near critical solutions.  We find that the gravitational radiations damp rapidly, suggesting that the unstable mode is not radiating.  The determination of the possible existence of a non-spherical unstable mode/bifurcation point (\cite{Choptuik03}) requires a resolution beyond the computational resources available to us.

The existence of critical collapses in compact objects formed in collisions of NSs modeled by a commonly used EOS is by itself interesting.  Further, the facts that (1) the growth timescale of the unstable mode is short ($0.054ms$), (2) it can appear in systems which deviate significantly from spherical, and (3) most importantly, a critical collapse could result through the softening of the EOS, suggest that critical collapses may actually occur in nature, e.g., a proto-NS initially supported by thermal pressure gradually cools down through neutrino emissions.  When the EOS gradually changes in the timescale of seconds, the proto-NS evolves from the sub-critical regime toward the critical regime.  The unstable mode, which grows in a timescale shorter than the evolution timescale of the EOS, may then be triggered, leading to a critical collapse.

\begin{figure*}
\hspace{-0.06in}
\begin{minipage}[t]{.3\textwidth}
\begin{center}
\includegraphics[angle=-90,width=2.2in]{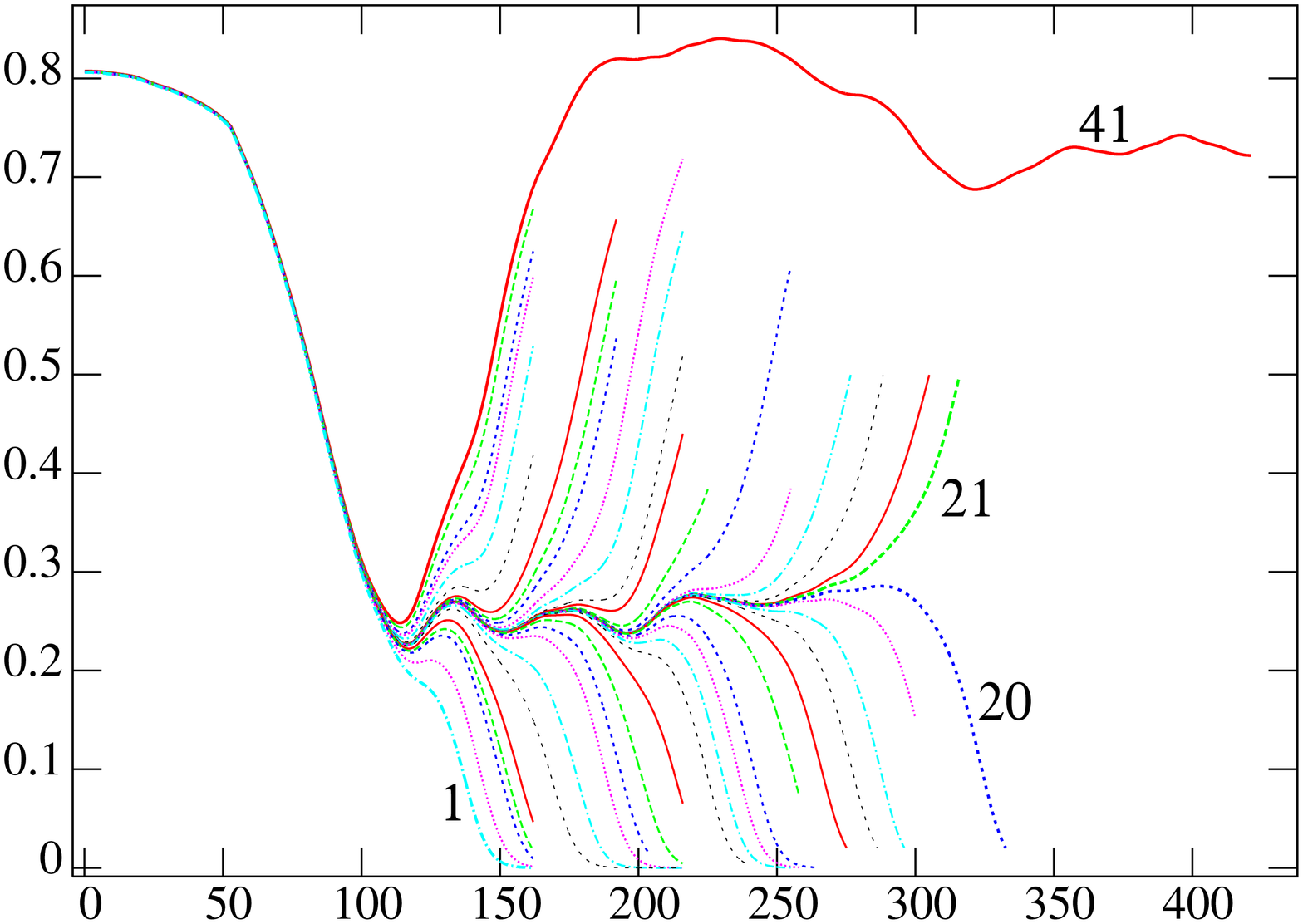}
\hfill
\vspace{-0.06in}
\caption{Lapse functions at center of collision vs time for NSs with slightly different masses.}  
\label{fig:headon_crit_lapse_d12}
\end{center}
\end{minipage}
\hspace{0.02\textwidth}
\begin{minipage}[t]{.3\textwidth}
\begin{center}
\includegraphics[angle=-90,width=2.2in]{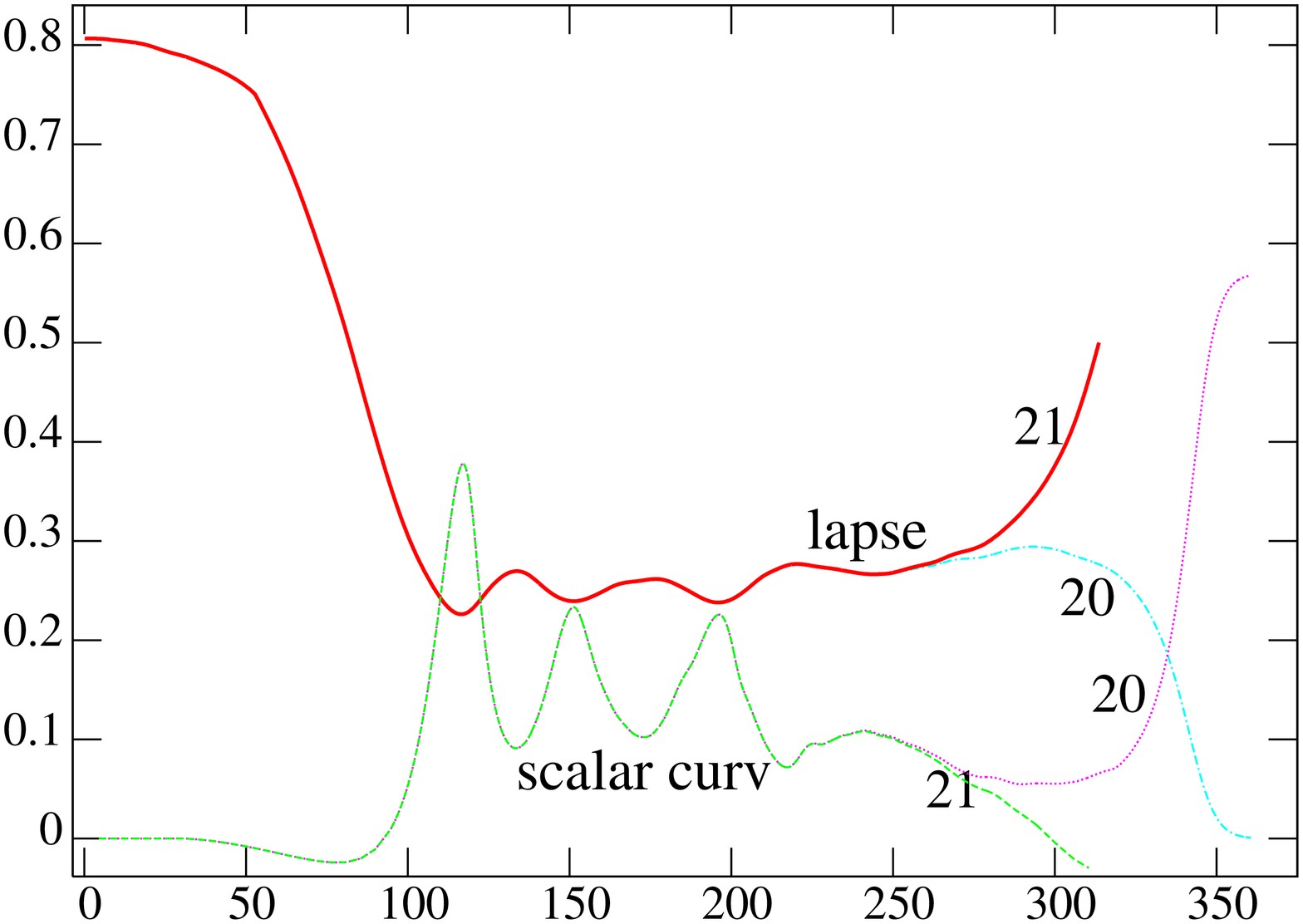} 
\hfill
\vspace{-0.06in}
\caption{Comparing the lapse functions and the 4-d scalar curvatures for the cases of lines 20 and 21 in Fig.~\ref{fig:headon_crit_lapse_d12}.}  
\label{fig:headon_crit_lapse_curv4_d12}
\end{center}
\end{minipage}
\hspace{0.02\textwidth}
\begin{minipage}[t]{.3\textwidth}
\begin{center}  
\includegraphics[angle=-90,width=2.2in]{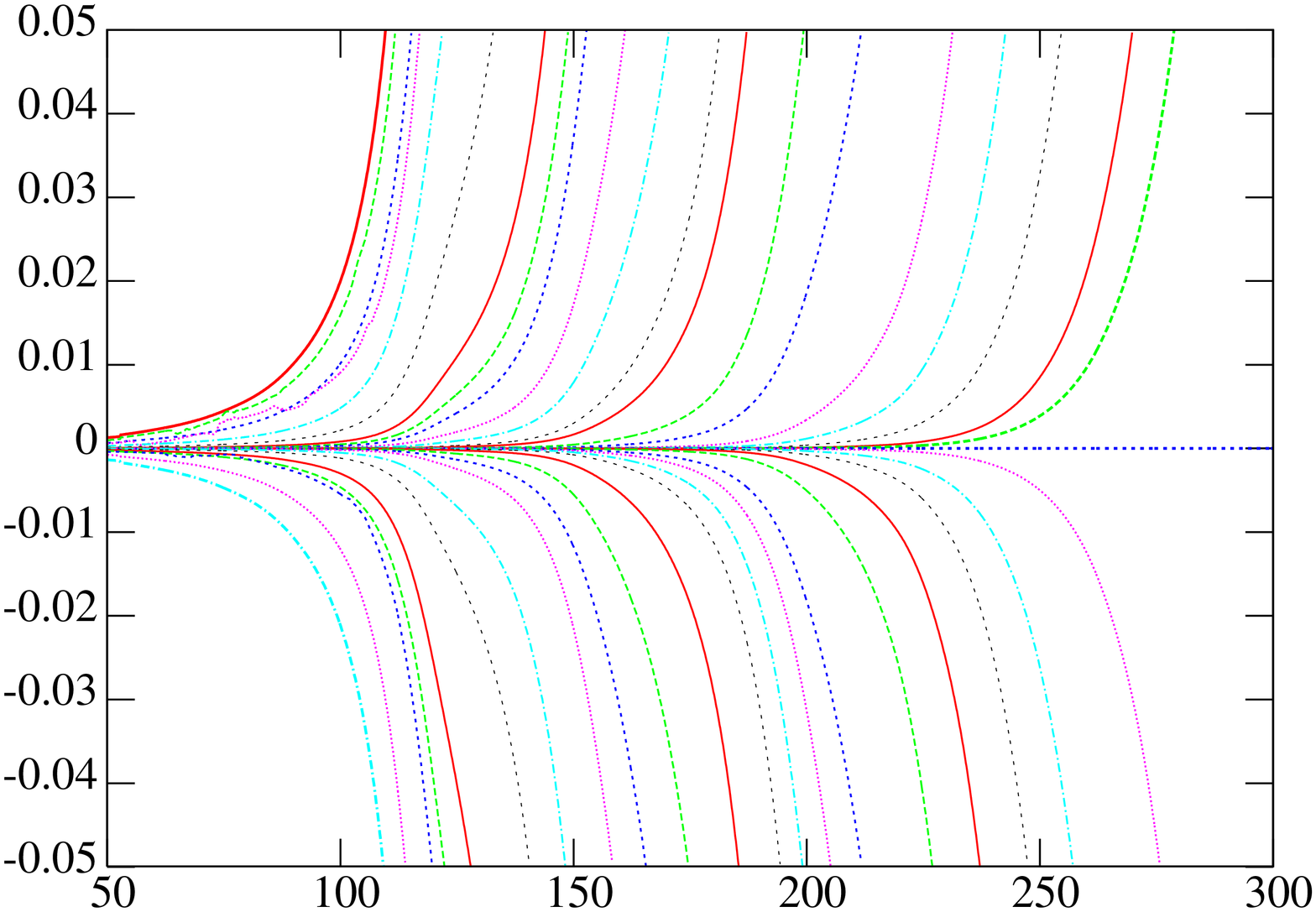}
\hfill
\vspace{-0.06in}
\caption{$(\alpha- \alpha _*)/\alpha_*$ v.s. coordinate time. }  
\label{fig:lapse_diff_05_d12}
\end{center}
\end{minipage}
\hfill

\vspace{-0.24in}
\end{figure*}

\paragraph*{\bf Sec. 2. GR-Astro-2D and the Numerical Setup.} 
Simulations of NS collisions are carried out using the GR-Astro code, which solves the coupled system of the Einstein equations and the general relativistic hydrodynamic (GR-Hydro) equations as described in \cite{Font98b,Font01b,Miller03b}.  The choice of numerical schemes as well as conventions used in this paper are given in \cite{Miller03b}.  A module GR-Astro-2D specially adapted to axisymmetric systems is constructed to provide the high resolution needed for this investigation.  GR-Astro-2D uses the same spacetime and GR-Hydro evolution routines as GR-Astro but applies symmetry conditions to restrict the simulation to a thin layer of spacetime ($5$ grid points across) containing the symmetry axis \cite{Alcubierre99a}.  Initial data sets for head-on collisions of NSs are constructed by solving the Hamiltonian and momentum constraints (HC and MCs) representing two NSs with equal mass separated by a distance along the $z$-axis and boosted towards one another at a prescribed speed.  The numerical evolutions in this paper are carried out with the $\Gamma$ freezing shift (or no shift) and the ``$1+\log$" lapse (for details of the shift and lapse conditions, see \cite{Miller03b}).  Extensive convergence tests have been performed including tests showing that the simulations are converging throughout the evolution.  In particular the critical index extracted (see below) has been shown to converge in a $1 ^{st}$ order manner, which is the designed rate of convergence due to our use of high resolution shock capturing treatment.

NSs in this paper are described by a polytropic EOS: $P = (\Gamma - 1) \rho \epsilon $ with $\Gamma = 2$ (and cases close to $2$).  Here $\rho$ is the proper rest mass density and $\epsilon$ is the proper specific internal energy density.  Notice that the "kinetic-energy-dominated" assumption has not been made, unlike earlier investigations of the critical collapses of perfect fluid systems (for review, see \cite{Neilsen98,Gundlach03,Wang01}). Initial data sets are constructed with $P = k \rho^\Gamma$, where $k = 0.0298 c^2/\rho_n$ ($\rho_n$ is the nuclear density, approximately $2.3$ x $10^{14} \; \rm{g/cm}^3$).  
For this EOS, the maximum stable NS configuration
has an Arnowitt-Deser-Misner (ADM) mass of $1.46 M_{\odot}$ and a baryonic mass of $ 1.61
M_{\odot}$.

\paragraph*{\bf Sec. 3. Existence of the Critical Phenomena in Head-on Collisions of NSs.} 

In the first set of simulations, the two NSs are initially at a fixed distance (the maximum density points of the two NSs are separated by $3R$, where $R \sim 9.1 M_\odot $ is the coordinate radius of the NSs).  The initial velocities of the NSs are that of freely falling from infinity, determined by the Newtonian formula plus the 1PN correction.  For the configurations investigated in Fig.~\ref{fig:headon_crit_lapse_d12} where the baryonic mass of each of the NS ranges from $0.786M_{\odot}$ to $0.793M_{\odot}$, the initial speed ranges from $0.15537$ to $0.15584$ (in units of $c=1$).  The computational grid has $323 \times 5 \times 323$ points, covering a computational domain of ($\pi r^2 \times$height $)=(\pi \times 38.5^2 \times 77.0 ){M_{\odot}}^3$.  Each NS radius is resolved with $76$ grid points, taking advantage of the octane- and axi-symmetry of the problem.

Fig.~\ref{fig:headon_crit_lapse_d12} shows the evolution of the lapse function $\alpha$ at the center of the collision as a function of the coordinate time, for systems with slightly different masses (all other parameters, including physical and numerical parameters, are the same). The line labeled 1 in Fig.~\ref{fig:headon_crit_lapse_d12} (which dips to $0$ near $t \sim 150 M_{\odot}$) represents the case of  $0.793 M_{\odot}$.  We see that after the collision, $\alpha$ promptly "collapses" to zero, signaling the formation of a black hole.  Note that the total baryonic mass of the merged object $1.59 M_{\odot}$ is {\it less} than the maximum stable mass of a TOV solution of the same EOS in equilibrium.  The prompt gravitational collapse of the merged object with such a mass indicates that it is in a state that is very different from being stationary \cite{Miller01a,EvansE02}.

The line labeled 41 in Fig.~\ref{fig:headon_crit_lapse_d12} (which rises at $t \sim 120 M_{\odot}$) represents the case where each of the NSs has the baryonic mass $0.786 M_{\odot}$.  The lapse at the collision center dips as the two stars merge, then rebounds.  The merged object does not collapse to a black hole but instead form a stable NS in axisymmetric oscillations.  The lapse at the center of the merged object oscillates around a value of $0.71$, with a period of about $160 M_{\odot}$.

\begin{figure*}
\hspace{-0.06in}
\begin{minipage}[t]{.3\textwidth}
\begin{center}
\includegraphics[angle=-90,width=2.2in]{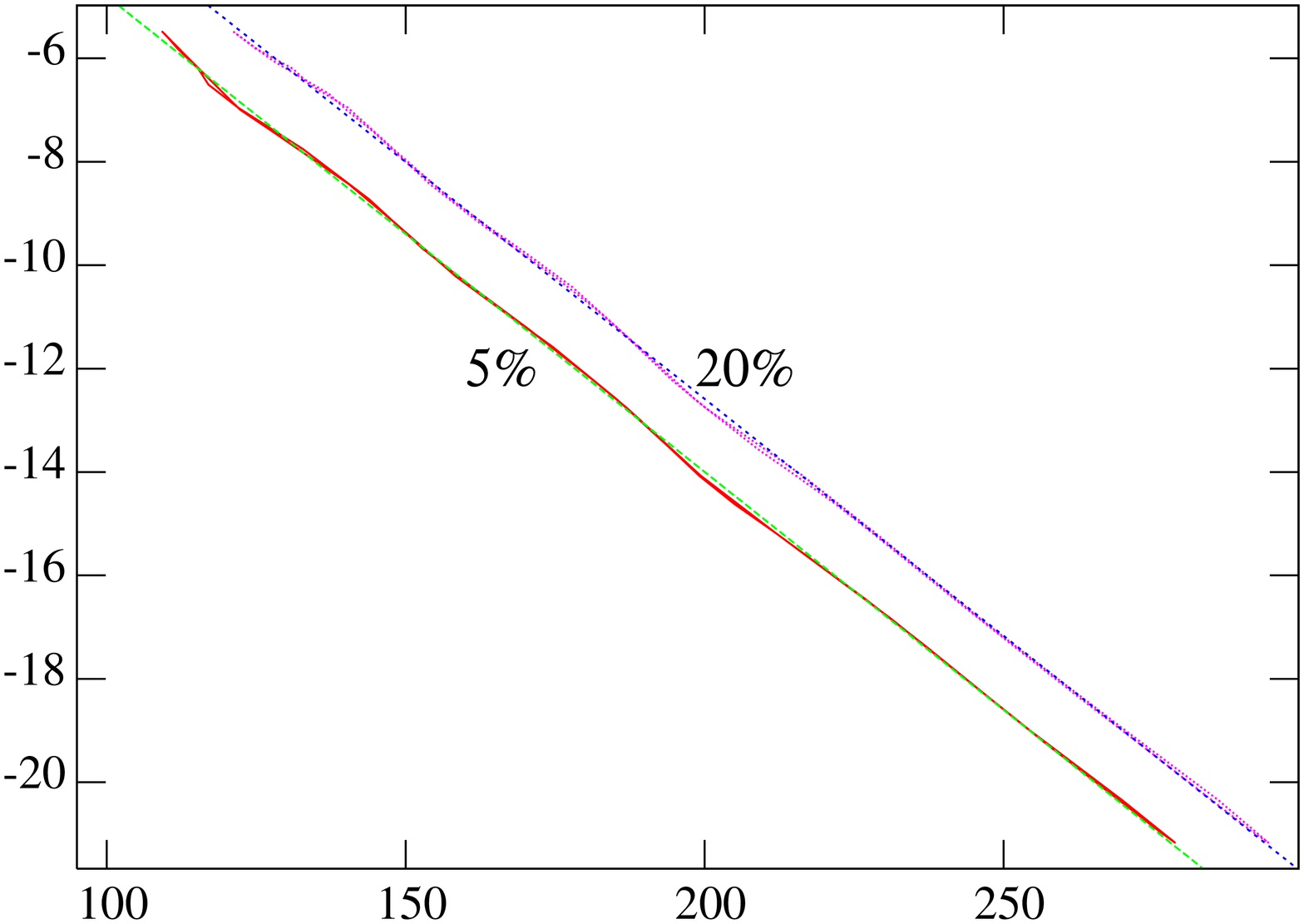}
\hfill
\vspace{-0.06in}
\caption{$Log(\rho_c-\rho_c^*)$ vs. the departure time determined with Fig.~\ref{fig:lapse_diff_05_d12}; the slope gives the critical index.}  
\label{fig:crit_calc_index_d12}
\end{center}
\end{minipage}
\hspace{0.02\textwidth}
\begin{minipage}[t]{.3\textwidth}
\begin{center}  
\includegraphics[angle=-90,width=2.2in]{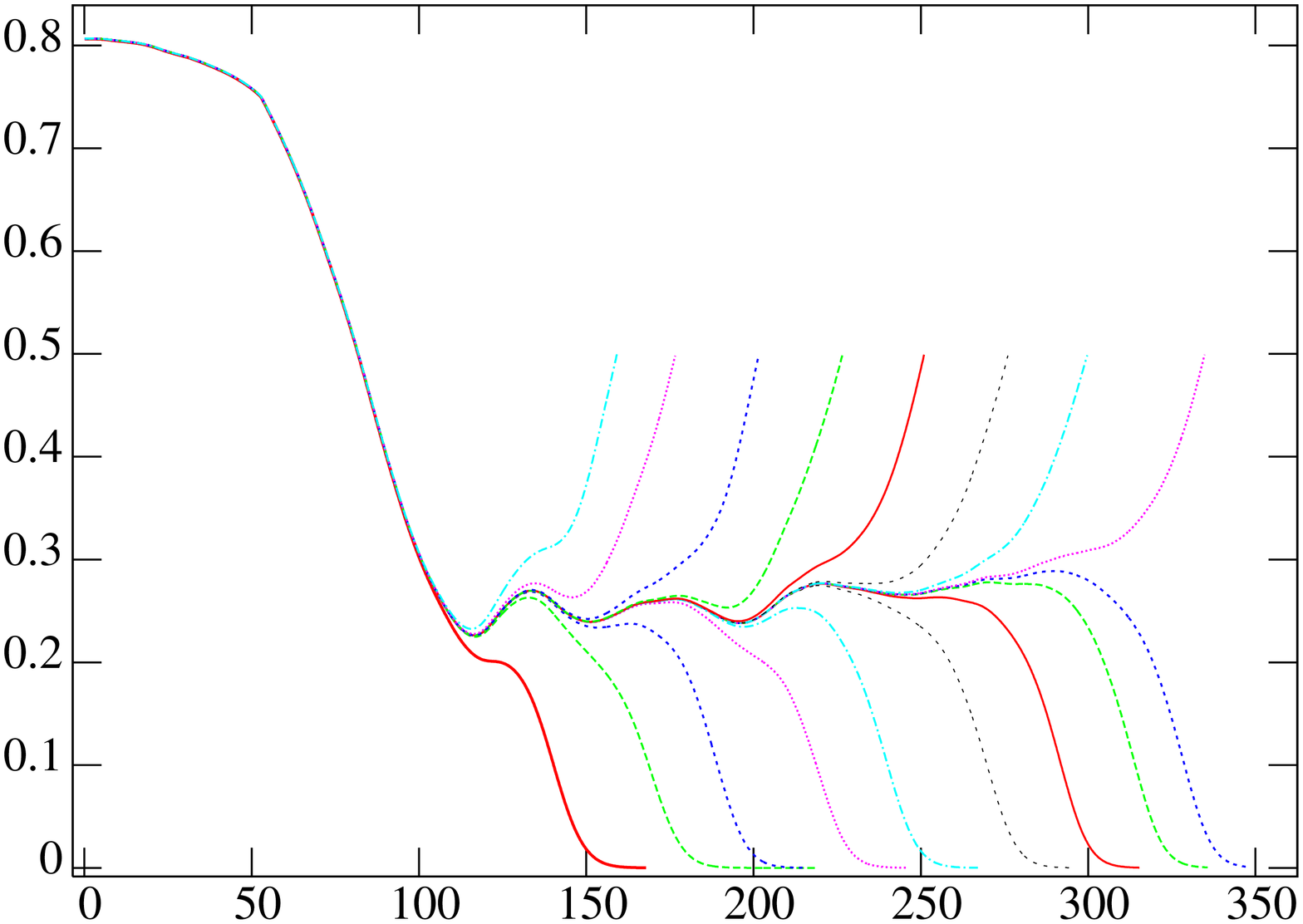}
\hfill
\vspace{-0.06in}
\caption{Lapses at the collision center for systems with the polytropic index $\Gamma$ varying between $1.9997$ and $2.0001$.}  
\label{fig:headon_crit_lapse_g12}
\end{center}
\end{minipage}
\hspace{0.02\textwidth}
\begin{minipage}[t]{.3\textwidth}
\begin{center}  
\includegraphics[angle=-90,width=2.2in]{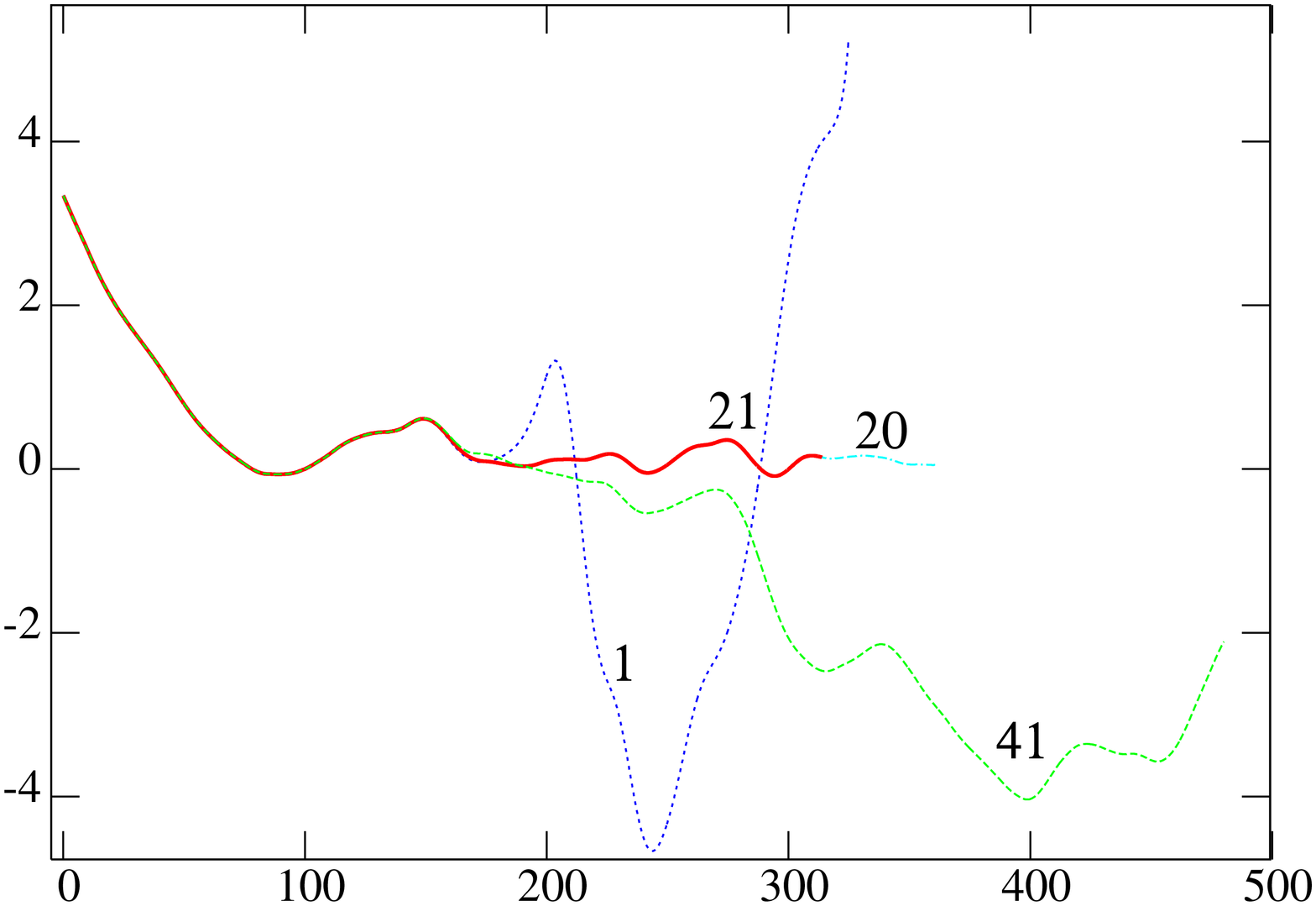} 
\hfill
\vspace{-0.06in}
\caption{The dominant l=2, m=0 wave components. Line numbers here correspond to those in Fig.~\ref{fig:headon_crit_lapse_d12}.}  
\label{fig:headon_crit_wave_d12}
\end{center}
\end{minipage}
\hfill

\vspace{-0.20in}
\end{figure*}

For configurations with masses between the bottom line(1) and top line (41), the
lapse $\alpha$ would rebound, dip etc., before eventually dipping to zero (a
black hole is formed) or going back up (a NS is formed).  The critical solution is found by fine tuning $\rho _ c$, the proper mass density as measured by an observer at rest with the fluid at the center of the star at the initial time. For the numerical setup used in the study, at around $\rho _ c = 6.128202618199\times 10^{-4}$ (mass of each NS $=0.79070949026 M_{\odot}$), a change of the $\rho _ c$ by the $10^{th}$ significant digit changes the dynamics from collapse to no collapse.  In Fig.~\ref{fig:headon_crit_lapse_d12} we see that for these near critical configurations $\alpha$  oscillates at about $0.255$ with a period of $\sim 40 M_{\odot}$.  As the lapse is given by the determinant of the 3 metric, this represents an oscillation of the 3 geometry.

For a more invariant measure, in Fig.~\ref{fig:headon_crit_lapse_curv4_d12} we plot as dotted and long dashed lines the 4-D scalar curvature $R$ at the collision center for two of the near critical solutions (lines $20$ and $21$ in Fig.~\ref{fig:headon_crit_lapse_d12}; they are the last ones to move away from the exact critical solution at $t \sim 300 M_{\odot}$).   We see that $R$ oscillates with the same period as the determinant of the 3 metric (the lapse).  As $\alpha$ collapses to zero, $R$ blows up and in each such case we find an apparent horizon, indicating the formation of a black hole.  Similar oscillatory behavior has been seen in other critical collapse studies \cite{Choptuik03,Abrahams94a,Gundlach03,Wang01}.  We note that at late time $R$ of the sub-critical case (line 21) tends to a small negative value as a static TOV star should.

We note that while in Fig.~\ref{fig:headon_crit_lapse_d12} a change in the $10^{th}$ significant digit of the total mass of the system can change the dynamics from that of sub-critical to supercritical, this does not imply that we have determined the critical point to the $10^{th}$ digit of accuracy.  The exact value of the critical point is affected by the resolution of the numerical grid as well as the size of the computational domain.  We have performed high resolution simulations with $76$ grid points per $R$, (with computational domain covering $8.5R$), and large computational domain simulations covering $34R$ (with resolution $38$ grid point per $R$).  Convergence tests in both the directions of resolution and size of the computational domain suggest that the total mass of the critical solution in the headon collision case with the EOS given is at $1.58 \pm 0.05 M_\odot$, with the error bound representing the truncation errors.

\paragraph*{\bf Sec. 4. Determination of the Critical Index.} 

The critical index $\gamma$ is determined through the relation $T = \gamma\log (p-p_*)$, where $T$ is the length of the coordinate time (which is asymptotically Minkowski) that a near critical solution with a parameter value $p$ stays near the exact critical solution with $ p_*$.  In Sec. 3 above, $p$ is taken as the central density $\rho_c$ of the initial NSs.  In Fig.~\ref{fig:lapse_diff_05_d12}, we plot $(\alpha- \alpha _*)/\alpha_*$ at the center of collision against the coordinate time, where $\alpha_*$ is the lapse of the critical solution to the best we can determine.  Only the last part of the evolution is shown.  We see explicitly the growth of the unstable mode driving the near critical solution away from the critical solution.  We defined the "departure time" $T_{0.05}$ as the coordinate time that a line in this figure reaches $\pm 0.05 =\pm 5\%$.  Likewise we define $T_{0.1}$, $T_{0.15}$ and $T_{0.2}$.  In Fig.~\ref{fig:crit_calc_index_d12}, the departure times $T_{0.05}$ and $T_{0.2}$ are plotted against the log difference of $p$ (taken to be $\rho_c$ as in Fig.~\ref{fig:headon_crit_lapse_d12}) between the near critical and the critical solutions.   With this, $\gamma_{\thinspace 0.05}$ defined as $T_{0.05} / \log (p-p_*)$ is found to be $10.87$, whereas $\gamma_{\thinspace 0.10}$, $\gamma_{\thinspace 0.15}$ and  $\gamma_{\thinspace 0.2}$ are found to be $10.92,10.93$ and $10.92$ respectively.   We see that the value of the critical index does not depend sensitively on the definition of the departure point.

\paragraph*{\bf Sec. 5. Universality.} 

The above study uses the total mass/central density of the initial NSs as the critical parameter $p$.  Next we fix the central density $\rho_c$ of the initial NSs at $6.12820305495\times 10^{-4} {M_\odot}^{-1}$.   The initial coordinate separation between the center of the two NSs is fixed to be $D=27.5M_\odot$. The initial velocity $v$ is taken to be the parameter $p$.  For each choice of $v$, the HC and MC equations are solved.  Convergence with respect to spatial resolutions and outer boundary location has been verified.  We find the same critical phenomena.  The critical index is extracted in the same manner and found to be $10.78 M_\odot $.
.   

Other choices of parameter $p$ have also been studied, including:(i) $p=D $, while fixing $\rho_c$ and $v$, and (ii) $p=\rho_c$ while fixing $v$ and $D$.   Note that the latter case is different from the case discussed in Secs. 3 and 4 above, where the initial velocity is determined by the free fall velocity up to the first PN correction.  In all cases studied, we see the same critical phenomena with consistent values of the critical index $\gamma$.

Next we ask: Is critical collapse possible only through fine tuning the initial data?   If true, we would not expect to see critical collapse phenomena in nature.   We investigate the possibility of taking $p= \Gamma$, the adiabatic index, as slow changes of the EOS could occur in many astrophysical situations, e.g., accreting NSs and during cooling of proto-NSs generated in supernovae.  We fix $D$, $\rho_c$, $v$ and vary $\Gamma$ away from $2$.  The evolution of the lapse at the center of collision is shown in Fig.~\ref{fig:headon_crit_lapse_g12}. We see behavior similar to that of Fig.~\ref{fig:headon_crit_lapse_d12}. 
The critical index $\gamma$ is found to be again $10.78 M_\odot$, consistent with the values found by fine tuning the initial configurations.  

\paragraph*{\bf Sec. 6. Gravitational Wave Signals From Near Critical Collapses.} 

Can near critical collapses have a signature in their gravitational radiations that one can identify in observation, in view of the possibility that near critical collapse of a neutron star like object may occur in nature?   

In Fig.~\ref{fig:headon_crit_wave_d12}, we show the dominant gravitational
wave component (even parity $L=2, m=0$ component of the Moncrief
gauge invariant $Q$ \cite{Moncrief74}, $z$-axis is the symmetry axis) for four cases shown in Fig.~\ref{fig:headon_crit_lapse_d12}.  For the line $41$, the merger leads to a NS oscillating in a non-spherical fashion and emits gravitational waves.  For the line $1$ the merged object moves away from the critical solution at around $t \sim 130$ and collapses to a black hole.  The asymmetric collapse generates gravitational wave.  The simulation was terminated at $t \sim 330$ as the metric functions can no longer be adequately resolved with the formation of the black hole.  In contrast, the radiations from the near critical solutions represented by lines $20$ and $21$ (the dotted-dashed line and the solid line respectively, nearly coinciding with one another all the way) decrease in amplitude.  At around $t \sim 270$, the unstable mode sets in and the two near critical solutions $20$ and $21$ move away from the critical solution (as shown in Fig.~\ref{fig:headon_crit_lapse_d12}).  However, there is only a negligible amount of wave emitted, independent of whether
the unstable mode leads to a black hole (line 20) or an oscillating NS 
(line 21).  This suggests that the unstable mode may be a spherical mode that does not radiate.  The waveforms reported here suffer from that the computational domain (covering only $\sim 1$ wavelength) cannot be extended further out, limited by the computational resources available to us. 
Further investigations of the waveforms will be reported elsewhere.

\paragraph*{\bf Sec. 7. Conclusion.} 
We found that the dividing line between prompt and delayed collapses
of a compact object formed in a head-on collision of NSs with
a polytropic EOS occurs at a mass less than the maximum stable mass of a single equilibrium star.  The EOS is one that is often used in modeling NSs.  There exists a type I critical collapse phenomena at the dividing line.   Universality of the phenomena with respect to different choices of the critical parameter is confirmed and the critical index extracted.  The growth time of the unstable mode which brings a
near critical solution away from the critical one is found to be $\sim 0.054ms$, for the EOS studied.  The study suggests that, despite the highly asymmetric initial data, the final critical collapse may not generate gravitational radiation.  

We found that critical collapses could happen without fine tuning of initial data, but instead, through a gradual change of the EOS.  This opens the intriguing possibility that, e.g., when a proto NS cools and loses
thermal support on a timescale longer than $0.054ms$, the collapse could exhibit critical behavior.   This and related questions will be investigated elsewhere.  

\acknowledgments

GR-Astro is written and supported by Ed Evans, Mark Miller, Jian Tao, Randy Wolfmeyer, Hui-Min Zhang and others.  GR-Astro-2D is developed and supported by K. J. Jin.  GR-Astro makes use of the Cactus Toolkit developed by Tom Goodale and the Cactus support group.  We thank Sai Iyer, Jian Tao, Malcolm Tobias, Randy Wolfmeyer, Hui-Min Zhang and other members of the WUGRAV group for discussions and support. The research is supported by the NSF Grant Phy 99-79985 (the KDI Astrophysics Simulation Collaboratory Project), NSF NRAC MCA93S025, and the McDonnell Center for Space Sciences at the Washington University.


\end{document}